\documentclass[12pt]{iopart}
\usepackage[dvips]{graphicx}
\def\spose#1{\hbox to 0pt{#1\hss}}
\def\lesssim{\mathrel{\spose{\lower 3pt\hbox{$\mathchar"218$}}
    \raise 2.0pt\hbox{$\mathchar"13C$}}}
\def\gtrsim{\mathrel{\spose{\lower 3pt\hbox{$\mathchar"218$}}
    \raise 2.0pt\hbox{$\mathchar"13E$}}}

\begin{document}

\title{Dark Matter Spikes and Indirect Detection}

\author{David Merritt}

\address{Rutgers University, New Brunswick, NJ 08903, USA}
\ead{merritt@physics.rutgers.edu}

\begin{abstract}
Annihilation radiation from supersymmetric particles 
at the Galactic center could be greatly enhanced
if the dark matter density is peaked 
around the supermassive black hole.
Arguments for and against the existence of
density spikes are reviewed.
Spikes are destroyed during mergers,
and there is strong evidence for this effect 
in {\it stellar} density profiles.
The dark matter spike at the Galactic
center probably suffered this fate.
\end{abstract}

Indirect detection schemes are based on searches
for gamma rays, neutrinos or other annihilation by-products
from supersymmetric particles in the dark matter halo of the 
Milky Way \cite{Feng01}.
The flux depends on the squared density of particles
integrated along the line of sight and the signal is greatly
enhanced in directions where the dark matter is clumped.
This includes the center of the Milky Way
where the density in a smooth halo would be 
maximum\cite{Berezinsky94,Bergstrom98}.
The signal from the Galactic center is further enhanced
if there is a dark matter ``spike'' associated with the
central supermassive black hole (SBH)\cite{Gondolo99}. 
Growth of a SBH causes the orbits of nearby
dark matter particles to shrink and their
density to rise; for a wide range of initial conditions,
the resulting profile is a steep power law, $\rho\sim r^{-2}$,
implying a formally divergent annihilation flux from 
near the SBH\cite{Gondolo99}.

This article reviews the arguments for and against the
existence of dark matter density spikes.
A useful guide to the distribution of dark matter
at the very centers of galaxies is the {\it stellar} 
distribution; in the absence of gaseous dissipation,
stars act like a nearly collisionless fluid 
and should react to the presence of a SBH in 
the same way as the dark matter.
Steep spikes in the stellar luminosity profile are only
seen in galaxies which show evidence of dissipative
formation.
Other galaxies exhibit shallow inner profiles,
and this is consistent with a model in which
binary SBHs injected energy into the stellar fluid during
the mergers that formed the galaxies.
Mergers almost certainly occurred during the formation
of SBHs and their host bulges, implying destruction
of the dark matter spikes.

\section{The Adiabatic Growth Model}

If a black hole grows at the center of a collisionless
fluid, the density around it also grows as the black hole's 
gravity causes the surrounding orbits to shrink.
The result is simplest to compute under the ``adiabatic''
approximation in which the growth time is long compared with
orbital periods; this is reasonable in the case of SBHs
at galactic centers, whose growth time is thought to
be $\sim 10^8$ yr compared with orbital periods of 
$\sim 10^6$ yr.
The resulting density profile depends somewhat on 
the initial state.
One possibility is a constant-density, isothermal core;
the resulting spike is a power law, $\rho\propto r^{-1.5}$,
$r\lesssim r_h \equiv GM_\bullet/\sigma^2$,
with $M_\bullet$ the final black hole mass and
$\sigma$ the 1D particle velocity dispersion before
the appearance of the black hole\cite{Peebles72,Young80}.


\begin{figure}
  \centering
    \includegraphics[width=2.5in]{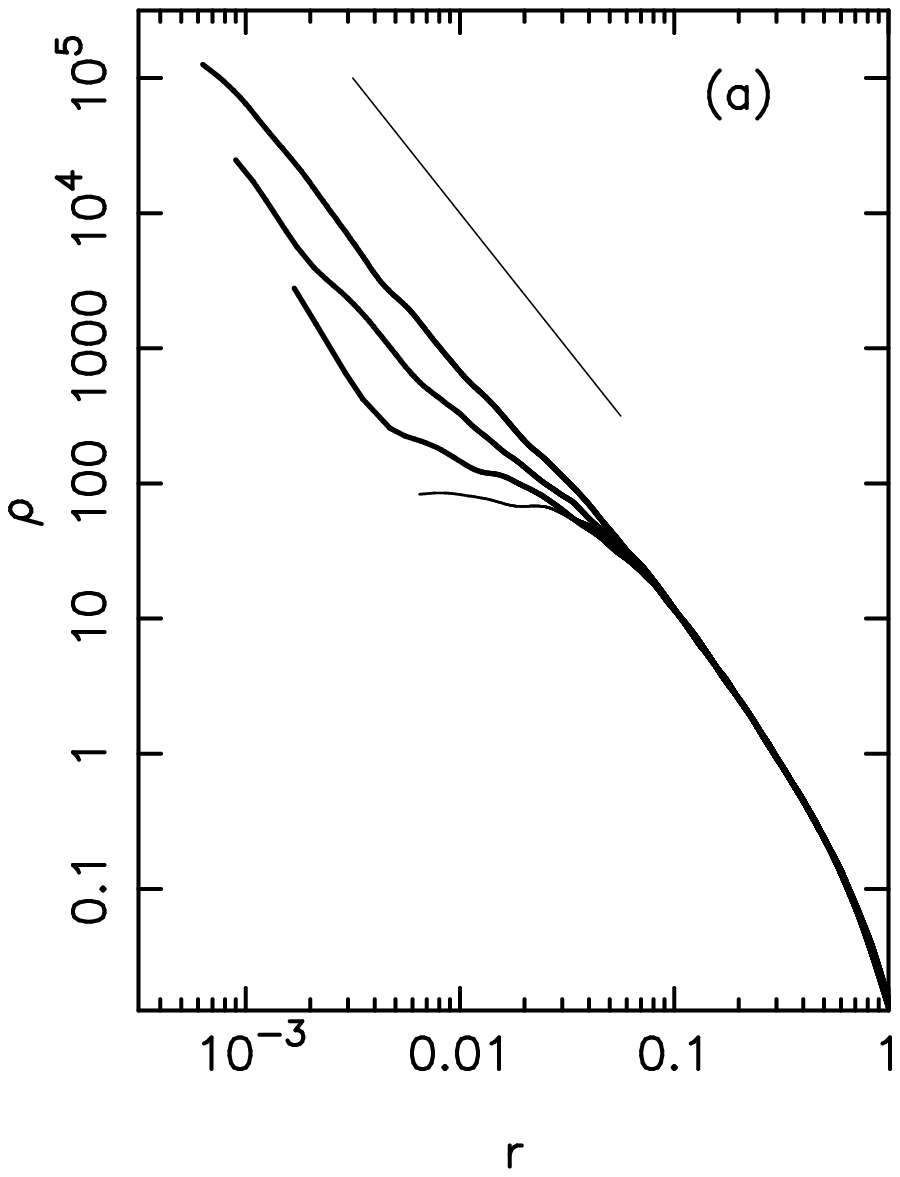}%
	\hspace{0.5in}%
    \includegraphics[width=2.5in]{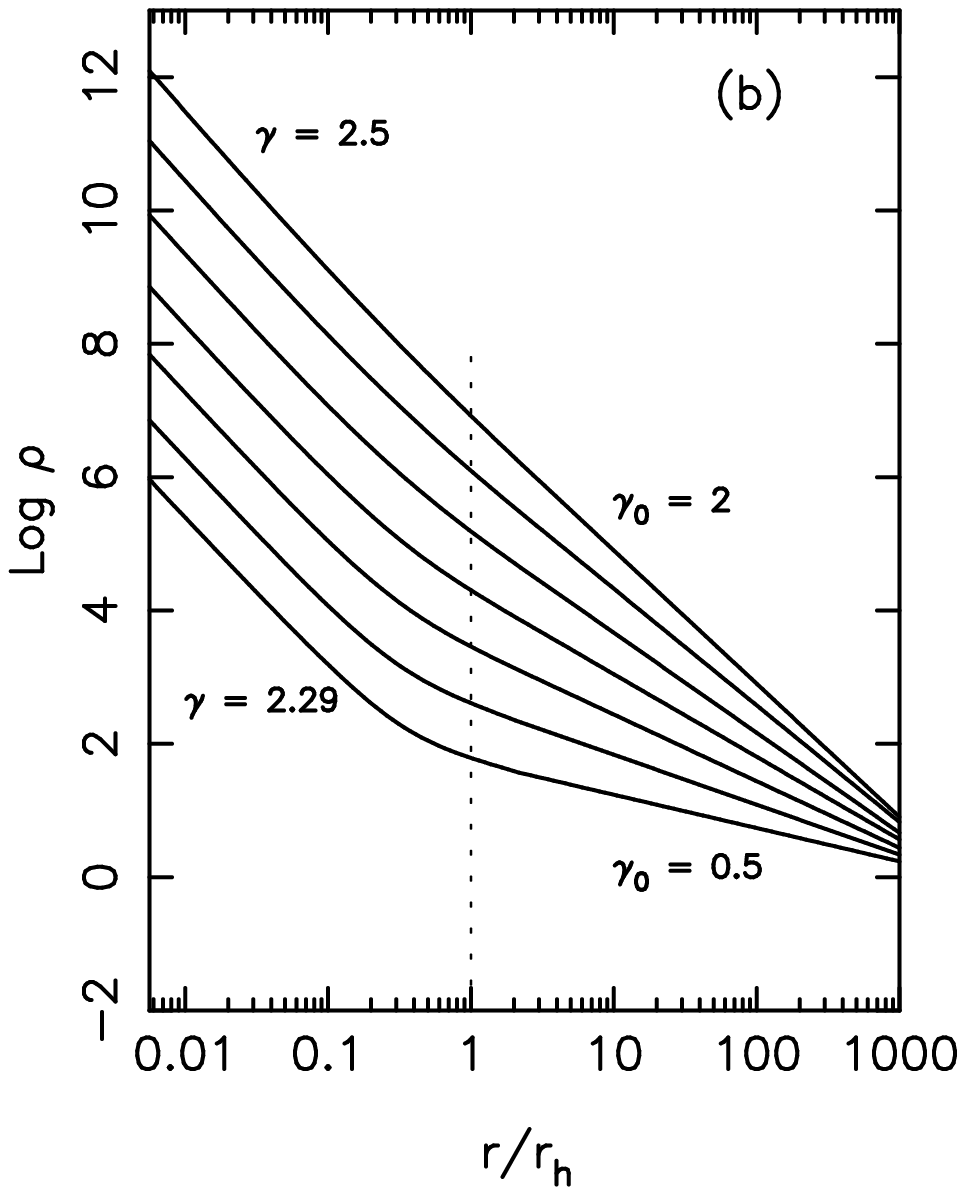}
\caption{Spike formation by adiabatic growth of a black hole.
(a) Thin curve is the initial model, formed by collapse
in an $N$-body simulation.
Heavy curves are density profiles after growth of a
central ``black hole'' containing $0.3\%$, $1\%$ and $3\%$
of the total mass.
The reference line has a logarithmic slope of $-2$.
(From Ref. 5.)
(b) Initial models were spherical isotropic halos
with power-law density profiles,
$\rho\propto r^{-\gamma_0}$;
$\gamma_0$ increases upwards in steps of $0.25$.
The radial scale is normalized to $r_h$ in the initial halo.
The slope of the final profile at $r\lesssim r_h$ is almost
independent of the initial slope.
}
\label{fig:grow}
\end{figure} 

A more reasonable guess for the initial profile is
a power law, $\rho\propto r^{-\gamma_0}$; power-law central profiles 
are generic outcomes of hierarchical structure
formation simulations\cite{NFW96,Moore99}, although the index
of the power law is debated.
Adiabatic growth in a power law gives a final
profile that is also a power law near the black hole
but with steeper index\cite{Gondolo99}:
\begin{equation}
\rho_f(r) \propto r^{-\gamma},\ \ \ \ \gamma=2 + {1\over 4-\gamma_0}\ \ \ \ (0<\gamma_0<2).
\end{equation}
The final index is almost independent of the initial index:
$2.25 < \gamma < 2.5$ (Figure 1).
Other reasonable initial profiles can be found\cite{Quinlan95}
that generate final profiles which fill the gap 
between the power laws and the isothermal sphere.
Hence in the absence of detailed knowledge about the initial state,
the adiabatic growth model predicts density profiles such that
\begin{equation}
\rho_f(r) \propto r^{-\gamma},\ \ \ \ 1.5\lesssim\gamma\lesssim 2.5.
\end{equation}
Even slopes at the low end of this range should not
be ruled out since the latest simulations\cite{Power03} suggest 
that $\gamma_0$ may decline monotonically toward the center.
Essentially nothing is known about the structure
of CDM halos on the very small ($\lesssim 100$ pc) scales
relevant to the formation of a spike\cite{Moore01}.

\section{Stellar Density Profiles}

The predictions of the adiabatic growth model can be tested against 
{\it stellar} luminosity profiles in galactic nuclei.
The time scales for both physical collisions and gravitational
near-encounters exceed a Hubble time for the stars in most 
nuclei\cite{Faber97}; 
in the absence of ongoing star formation,
nuclei should have retained whatever density profiles 
were set up when the SBH gained its current size.
Observed profiles are indeed well described as power laws
within the SBH's sphere of influence,
but with slopes that vary systematically as a function of
galaxy luminosity\cite{Merritt96,Gebhardt96}.
Faint ellipticals and bulges, $M_V\gtrsim -20$,
exhibit roughly the range of
slopes predicted by the adiabatic growth model,
$1.5\lesssim\gamma\lesssim 2.5$.
However bright galaxies have shallower inner profiles,
$0\lesssim\gamma\lesssim 1.5$.

There are problems with identifying even the steep
profiles in faint galaxies with the adiabatic growth 
picture\cite{Merritt03}.
Spikes produced by adiabatic growth almost always exhibit 
an inflection at $r\sim r_h$ (Figure 1); the logarithmic
slope increases inwards.
In real galaxies, the slope {\it decreases} inwards;
inflections are virtually never seen, except at much smaller 
radii where they are associated with point-like or non-thermal
nuclei.
Avoiding an inflection requires fine-tuning of the initial
conditions: either the initial profile has to be steep,
$\gamma_0\approx 2$, which obviates the need for adiabatic
growth; or the SBH has to be just the right size that
the pre-existing slope at $r\sim r_h$ matches the
final slope $\gamma$.

The rapid rotation, high densities and disky isophotes of
low-luminosity ellipticals and bulges imply that gas dynamics played
a dominant role in their formation\cite{Bender97}.
The steep stellar density spike at the
center of the Milky Way, $\rho_*\sim r^{-2}$,
may be the result of sustained star 
formation from a reservoir of dense molecular gas\cite{Serabyn95}.
Semi-analytic models of galaxy formation\cite{Haehnelt00}
suggest that galaxies fainter than
$M_V\sim -20$ were formed from gas-rich progenitors. 
These arguments suggest that the structure of nuclei in 
faint elliptical galaxies and bulges is a combined result of SBH growth 
and gas dynamics.
For this reason, faint galaxies are not ideal
testing grounds for the adiabatic growth model.

A cleaner test of the model comes from bright elliptical galaxies, 
and here the model clearly fails:
nuclear profiles in bright galaxies are almost always flatter than 
the minimum slope allowed by adiabatic growth,
$\gamma\sim 1.5$.
Profiles are sometimes so flat that they
are best described as ``cores,'' regions of nearly-constant
density; indeed central {\it minima} have been claimed
in a handful of galaxies\cite{Lauer02}.
Core radii can be $\sim 10^2$ pc or more.
Something other than adiabatic SBH growth 
is required to explain this structure.

\section{The Binary Black Hole Model}

A natural way to explain the low central densities
of bright galaxies is mergers.
A SBH is effective at tidally disrupting the 
density cusp of an infalling galaxy\cite{Merritt01,Merritt02}.
If the latter also contains a SBH, a binary SBH will
form, which ejects passing stars or dark matter
particles via the ``gravitational 
slingshot.''\cite{Saslaw74,Quinlan96}
A binary SBH ejects of order its own mass
before either coalescing in a burst of gravitational
radiation, or, if the supply of ejectable matter runs out,
stalling at a separation of $\sim 0.01-1$ pc\cite{MM01}.

\begin{figure}
  \resizebox{\hsize}{!}{
    \includegraphics{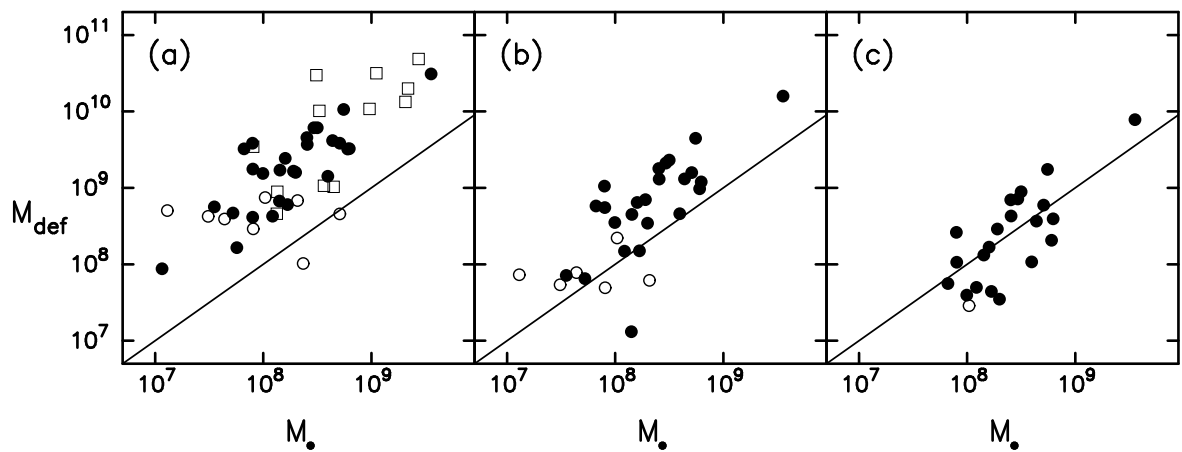}}
\caption{Mass deficit vs SBH mass for three different assumed
values of $\gamma_0$, the assumed logarithmic slope of the density
cusp before changes due to ejection of stars by a binary
SBH. (a) $\gamma_0=2$; (b) $\gamma_0=1.75$; (c) $\gamma_0=1.5$.
Solid lines are $M_{def}=M_\bullet$; units are solar masses. 
(From Ref. 26.)
}
\label{fig:mdef}
\end{figure}

The binary black hole model does a credible job of explaining
the central density profiles of bright galaxies.
Figure 2 plots the {\it mass deficit}\cite{Milos02}:
\begin{equation}
M_{def} \equiv 4\pi\int_0^{r_b}\left[\rho_*(r_b)\left({r\over r_b}\right)^{-\gamma_0} - \rho_*(r)\right]r^2dr,
\end{equation}
the difference in the integrated mass between the observed
density profile, and a $\rho_*(r)\propto r^{-\gamma_0}$ 
profile extrapolated inward from the turnover radius $r_b$.
There is a good correlation of $M_{def}$ with $M_\bullet$,
although the constant of proportionality can lie anywhere between 
$\sim 1$ and $\sim 10$ depending on the choice of $\gamma_0$.
Larger ratios would be consistent with a picture in which
galaxies form through a succession of mergers\cite{MM01}.

Interestingly, another class of stellar system exhibits 
both spikes and cores: the globular clusters.
Globular clusters may contain ``intermediate mass''
($M_\bullet\sim 10^3 M_\odot$) black holes,
although the evidence\cite{Gebhardt02} is not compelling.
However they almost certainly contained an early population
of $\sim 10 M_\odot$ black holes, remnants of the first
generation of stars\cite{SPZ00}, which would have spiralled to the center
and displaced the lighter stars before ejecting themselves 
via the gravitational slingshot.
This mechanism appears capable of creating the cores\cite{MM02}.
The steep central spikes seen in some globular clusters\cite{Djorgovski86}
are probably a consequence of collisional relaxation.

\section{Implications for Dark Matter Spikes}

A dark matter spike could have survived at the center
of the Milky Way if no mergers occurred
since the era of SBH formation\cite{Bertone02}.
But the formation of the SBH was itself probably
triggered by a major merger which channeled gas
into the central regions\cite{Haehnelt00}; indeed a binary SBH may
have played a crucial role in the channeling\cite{Gould00}.
It is hard to see how a dark matter spike could have survived
such an event, and in fact the quasi-stationary conditions
critical for the formation of the spike may never have 
existed\cite{Ullio01}.
These arguments suggest that a dark matter spike is unlikely
at the center of the Galaxy,
even if the Milky Way {\it disk} has not suffered
a major merger in the last $\sim 12$ Gyr\cite{Wyse01}.
A conservative upper limit
on the density of neutralinos at the Galactic center 
would be that given by CDM models of structure formation,
ignoring the SBHs\cite{Merritt02}.
Spike-free density profiles might still allow testing of
significant portions of MSSM parameter space\cite{Ullio01b}.

\section*{Acknowledgments}
This work was supported by NSF grants AST 00-71099 and
AST 02-06031,
by NASA grants NAG5-6037 and NAG5-9046, and by
grant HST-AR-08759 from STScI.

\end{document}